\documentclass[12pt]{elsarticle}

\usepackage{graphicx}
\usepackage{amsmath}
\usepackage{amsthm}
\usepackage{amsfonts}
\usepackage[brazil]{babel}
\usepackage[utf8]{inputenc}

\usepackage{amsthm}

\def\|{\' \i }

\newcounter{bla}

\journal{Yet to be decided}

\newtheorem{teor}{Theorem}[section]

\newtheorem{obs}{Remark}[section]
\newtheorem{defin}{Definition}[section]

\newtheorem{algor}{Algorithm}[section]

\begin{document}

\title{{A generalization of the $S$-function method applied to a Duffing-Van der Pol forced oscillator}}

%\author[uerj]{M.S. Cardoso}
%\ead{mscardoso@gmail.com.br}

\author[uerj]{L.G.S. Duarte}
\ead{lgsduarte@gmail.com.br}

\author[uerj]{L.A.C.P. da Mota\corref{cor1}}
\ead{lacpdamota@uerj.br or lacpdamota@dft.if.uerj.br}

\cortext[cor1]{Corresponding author {\footnotesize \newline L.G.S. Duarte and L.A.C.P. da Mota wish to thank Funda\c c\~ao de Amparo \`a Pesquisa do Estado do Rio de Janeiro (FAPERJ) for a Research Grant.}}

\address[uerj]{Universidade do Estado do Rio de Janeiro,
{\it Instituto de F\'{\i}sica, Depto. de F\'{\i}sica Te\'orica},
{\it 20559-900 Rio de Janeiro -- RJ, Brazil}}

\begin{abstract}
In \cite{Noscpcnv2016,Nosarxiv2017} we have developed a method (we call it the $S$-function method) that is successful in treating certain classes of rational second order ordinary differential equations (rational 2ODEs) that are particularly `resistant' to canonical Lie methods and to Darbouxian approaches. In this present paper, we generalize the $S$-function method making it capable of dealing with a class of elementary 2ODEs presenting elementary functions. Then, we apply this method to a Duffing-Van der Pol forced oscillator, obtaining an entire class of first integrals.
\end{abstract}


\maketitle
\section{Introduction}
\label{intro}

The most general methods for finding closed-form solutions and global first integrals of dynamical systems derive from Lie \cite{Lie} and Darboux \cite{Dar} approaches. Thus, it is unnecessary to stress the importance of procedures that compute Lie symmetries (symmetry continuous groups) or Darboux polynomials (that define algebraic invariant curves or surfaces) associated with a given differential equation (DE) or a system of DEs. However, even these powerful approaches sometimes fail (this usually happens in cases where the Darboux polynomials have a relatively high degree, say $\ge$ 4, and / or the Lie symmetries are dynamical or nonlocal). Therefore, there is a constant search for generalizations of these methods, which, in turn, leads to the definition of a series of concepts and alternative procedures to attack these more complicated cases. Among these, we highlight: \cite{AbrGovLea,GovLea,MurRom,MurRom2,Nuc2,MurRom3,GanBruSen,BluDri,MurRom4,MurRomRui} in the `symmetry  side' and \cite{Nosjpa2001,Nosjpa2002-2,royal,Noscpc2007,Nosjmp2009,FerGia,ChrLliPanWal,AcoLazMorPan,Zha,BosCheCluWei} in the `Darbouxian side'.

In \cite{Noscpcnv2016,Nosarxiv2017} we  have developed an algorithm to find Liouvillian first integrals of rational second order ordinary differential equations (rational 2ODEs). This procedure is based on the computation of the $S$-function\footnote{The $S$-function was created in \cite{Nosjpa2001} where we generalize the Prelle and Singer method \cite{PreSin} to deal with rational 2ODEs.} that is closely linked to the $\lambda$-symmetry concept created by Muriel and Romero in \cite{MurRom}. In this paper we generalize the method developed in \cite{Noscpcnv2016,Nosarxiv2017} and use it to find first integrals of a forced Duffing-Van der Pol oscillator \cite{Duf,Pol}. As the determination of the $S$-function is based on a Darbouxian type technique, we can do, during the search process, an integrability analysis in the same way we did in \cite{Nosjmp2009}.

In the first section we will briefly review the algorithm described in \cite{Nosarxiv2017}. Next, we present a generalization of the method that allows us to use the algorithm to deal with 2ODEs presenting elementary functions. In section 3 we use the method to make an analysis of the integrability region of the parameters present in a forced Duffing-Van der Pol oscillator. Finally we present our conclusions and ways to extend our work.

%%%%%%%%%%%%%%%%%%%%%%%%%%%%%%%%%%%%%%%%%%%%%%%%%%%%%%%%%%%%%%%%%%%%%
%%%%%%%%%%%%%%%%%%%%%%%%%%%%%%%%%%%%%%%%%%%%%%%%%%%%%%%%%%%%%%%%%%%%%

\section{The $S$-function method}
\label{sfunc}

In this section, we will introduce some basic concepts involving rational 2ODEs and some results (for the proofs see \cite{Nosarxiv2017}) that allowed for the construction of an algorithm to search for Liouvillian first integrals.

\subsection{The basic concepts and results}

Consider the rational 2ODE given by
\begin{equation}
\label{2oder1}
z'=\frac{dz}{dx}=\phi(x,y,z)=\frac{M(x,y,z)}{N(x,y,z)},  \,\,(z \equiv y'),
\end{equation}
where $M$ and $N$ are coprime polynomials in $(x,y,z)$.

\begin{defin}
A function $I(x,y,z)$ is called {\bf first integral} of the 2ODE {\em (\ref{2oder1})} if $I$ is constant over the solutions of {\em (\ref{2oder1})}.
\end{defin}

\begin{defin}
Let $I$ be a first integral of the {\em 2ODE (\ref{2oder1})}. The function defined by $S := I_y/I_z$ is called a \mbox{\boldmath $S$}{\bf -function} associated with the {\em 2ODE} through the first integral $I$.
\end{defin}

\begin{teor}
\label{teqS}
Let $S$ be a $S$-function associated with the {\em 2ODE} $\,z'=\phi(x,y,z)$. Then $S$ obeys the following equation:
\begin{equation}
\label{eqS}
D_x[S]=S^2+\phi_{z}\,S-\phi_y\,.
\end{equation}
\end{teor}

\begin{defin}
\label{1odeass}
Let $I$ be a first integral of the {\em 2ODE (\ref{2oder1})} and let $S(x,y,z)$ be the $S$-function associated with {\em (\ref{2oder1})} through $I$. The first order ordinary differential equation defined by
\begin{equation}
\label{1odeassdefs}
\frac{dz}{dy}= - S(x,y,z),
\end{equation}
where $x$ is taken as a parameter, is called {\bf 1ODE associated} with the {\em 2ODE (\ref{2oder1})} through $I$.
\end{defin}

\begin{teor}
\label{sol1odeass}
Let $I$ be a first integral of the {\em 2ODE (\ref{2oder1})} and let $S(x,y,z)$ be the $S$-function associated with {\em (\ref{2oder1})} through $I$. Then $I(x,y,z)=C$ is a general solution of the {\em 1ODE associated} with {\em (\ref{2oder1})} through $I$.
\end{teor}

\begin{obs}
\label{solnotinv}
Note that Theorem {\em \ref{sol1odeass}} does not imply that, if we solve the {\em 1ODE} associated {\em (\ref{1odeassdefs})}, we would obtain $I(x,y,z)=C$. The reason is that the variable $x$ (the independent variable of the {\em 2ODE (\ref{2oder1})}) is just a parameter in the {\em 1ODE (\ref{1odeassdefs})}.
\end{obs}
\begin{obs}
\label{solrelati}
Since any function of $x$ is an invariant for the operator  $\partial_y - S\,\partial_z$, i.e., $(\partial_y - S\,\partial_z)[F(x)] = 0$, the relation between a general solution $H(x,y,z)=K$ of the {\em 1ODE (\ref{1odeassdefs})} and the first integral $I(x,y,z)$ of the {\em 2ODE (\ref{2oder1})} is given by $\,I(x,y,z)={\cal F}\left(x,H\right)$, such that the function $\,{\cal F}\,$ satisfies
\begin{equation}
\label{calFeq}
D_x[I] = \frac{\partial {\cal F}}{\partial x} +\left(\frac{\partial H}{\partial x} + z\,\frac{\partial H}{\partial y} + \phi\,\frac{\partial H}{\partial z}\right)  \frac{\partial {\cal F}}{\partial H} = 0.
\end{equation}
\end{obs}

\bigskip

%%%%%%%%%%%%%%%%%
\subsection{Construction of a method}
\label{coam}
\hspace\parindent

If the 2ODE (\ref{2oder1}) presents a Liouvillian first integral $I$ such that the associated $S$-function is given by $\,S = \frac{P}{N}\,$ then, substituting it into the 1PDE $\,D_x[S]=S^2+\phi_{z}\,S-\phi_y\,$ we obtain:
\begin{equation}
\label{eqP}
-{P}^{2} - \left( N_x+z\,N_y+M_z \right) P+ D[P] -M\,N_y+M_y\,N=0,
\end{equation}
where $\,D \equiv N\,D_x = N\,\partial_x +z\,N\,\partial_y+M\,\partial_z$. Then we construct a polynomial ${\cal P}$ with undetermined coefficients and substitute it in (\ref{eqP}). After, we collect the resulting polynomial equation in the variables $x,\,y,\,z$ and equate the coefficients of each monomial to zero, obtaining a system of equations. If this system presents a solution we will have found $S$ and we can construct the associated 1ODE (and try to solve it). From the solution of the associated 1ODE we can find the first integral $I(x,y,z)$ by solving the characteristic system of de 1PDE (\ref{calFeq}). So, we can propose the following algorithm:

\bigskip

\begin{algor}  \
\label{as1}
	\begin{enumerate}
		\item Let $n_{max}$ =  $\max(degree (M)-1,degree (N))$.
		\item Let $n=0$.
		\item Let $n=n+1$.
		\item if $n>n_{max}$ then FAIL.
             \item Construct the $D_x$ operator.
		\item Construct a generic polynomial ${\cal P}$ of degree $n$ in $(x,y,z)$ with undetermined coefficients $a_i$.
		\item Substitute ${\cal P}$ in equation (\ref{eqP}), collect the resulting polynomial equation in the variables $x,\,y,\,z$ and equate the coefficients of each monomial to zero, obtaining a system ${\cal A}$ of algebraic equations.
		\item Solve the system ${\cal A}$ with respect to $\{a_i\}$. If no solution is found, then go to step 3.
		\item Substitute the solution in ${\cal P}$ (obtaining $P$) and construct $S = P/N$.
		\item Construct the associated 1ODE ($\frac{dz}{dy}=-\frac{P}{N}$) and try to solve it to obtain a solution $H(x,y,z)=C$. If no solution is found, then go to step 3.
		\item Solve $h=H(x,y,z)$ for one of the variables $x$, $y$ or $z$ (or any operand belonging to $H$ and to $D_x[H]$) and substitute it in the 1PDE (\ref{calFeq}). Then, try to solve the characteristic equation -- a 1ODE in $(x,h)$ -- to obtain ${\cal F}(x,h)=K$. If no solution is found, then go to step 3.
		\item Construct the first integral $I={\cal F}(x,H)$.
	\end{enumerate}
\end{algor}

\bigskip

%%%%%%%%%%%%%%%%%%%%%%%%%%%%%%%%%%%%%%%%%%%%%%%%%%%%%%%
%%%%%%%%%%%%%%%%%%%%%%%%%%%%%%%%%%%%%%%%%%%%%%%%%%%%%%%

\section{A generalization of the $S$-function method to deal with 2ODEs presenting elementary functions}
\label{gsfunc}

The main idea is to consider 2ODEs of the form
\begin{equation}
\label{2odefx}
z'=\frac{dz}{dx}=\phi(f(x),y,z)=\frac{M(g(x),y,z)}{N(g(x),y,z)},  \,\,(z \equiv y'),
\end{equation}
where $M$ and $N$ are coprime polynomials in $(g(x),y,z)$. If we perform a change of variables such that $u = f(x)$ into the 2ODE (\ref{2odefx}) we have
\begin{eqnarray}
\label{2odetrans1}
\frac{dy}{dx} &=& \frac{dy}{du}\frac{du}{dx} = \frac{dy}{du}\, f'(x), \\[4mm]
\frac{d^2y}{dx^2} &=& \frac{d}{dx}\left(\frac{dy}{dx}\right) =\frac{d}{dx} \left(\frac{dy}{du}\, f'(x) \right) = \nonumber \\[2mm]
                              &=& f''(x)\,\frac{dy}{du}+f'(x)^2 \, \frac{d^2y}{du^2}. \label{2odetrans2}
\end{eqnarray}
Substituting in the 2ODE (\ref{2odefx}) we obtain
\begin{equation}
\label{2odefu}
f''(x)\,\frac{dy}{du}+f'(x)^2 \, \frac{d^2y}{du^2}=\frac{M(g(x),y,\frac{dy}{du}\,f'(x))}{N(g(x),y,\frac{dy}{du}\,f'(x))}.
\end{equation}

\noindent
The idea is that if the function $f(x)$ is such that $g(x)=h_0(u)$ and $f'(x)=h_1(u)$, where $h_0$ and $h_1$ are rational functions of $u$, it is clear that the 2ODE (\ref{2odefu}) will be rational (for in that case, $f''(x)$ will also be a rational function of $u$). In this way, we can apply the $S$-funtion method to it. We can see that if $g(x)$ is an elementary function, this will work in many cases. For example, let $g(x)=(1+{\rm e}^{2\,x})$: in this case choosing $f(x)={\rm e}^{x}$ we have $g(x)=1+u^2$, $f'(x)=u$, $f''(x)=u$.

%%%%%%%%%%%%%%%%%%%%%%%%%%%%%%%%%%%%%%%%%%%%%%%%%%%%%%%
%%%%%%%%%%%%%%%%%%%%%%%%%%%%%%%%%%%%%%%%%%%%%%%%%%%%%%%

\section{A first integral for a forced Duffing Van der Pol oscillator}
\label{DVdP}

The Duffing–Van der Pol’s differential equation \cite{Duf,Pol} is an important mathematical model for representing dynamical systems having a single unstable
fixed point and a single stable limit cycle. This kind of equation represents many physical phenomena and, like many others nonlinear problems, the forced Duffing-Van der Pol oscillator does not have global analytical first integrals for all values of the parameters \cite{KSBRD,KyzOkn}. Therefore, it is very important to have tools that can analyse the integrability range of parameters' values. This has already been done for the force free Duffing Van der Pol oscillator \cite{ChaSenLak,GaoFen}. With the generalization just developed, we can apply the $S$-function method to achieve this purpose. Consider the following forced Duffing-Van der Pol oscillator DE:
\begin{equation}
\label{dvpeq}
\frac {d^{2}x}{d{t}^{2}} -\mu\, \left( 1- x^{2} \right) {\frac {dx}{dt}}+\alpha\,x +\beta\,x^{3}=f\cos \left( \omega\,t \right),
\end{equation}
where $x$ represents the displacement from the equilibrium, $\mu > 0$ is the damping parameter and $f\,\cos(\omega\,t)$ is the periodic driving function of time.

\bigskip

For this 2ODE, we can see that $g(t)=\cos(\omega\,t)$. So, if we choose $\tau=f(t)={\rm e}^{i\omega\,t}$, we will have $f'(t)=i\omega\,{\rm e}^{i\omega\,t}=i\omega\,\tau$,  $f''(t)=-\omega^2\,{\rm e}^{i\omega\,t}=-\omega^2\tau$ and $g(t)=\cos(\omega\,t)=(\tau+1/\tau)/2$. The transformation given by the eqs. (\ref{2odetrans1},\ref{2odetrans2}) is
\begin{eqnarray}
\label{2odet1}
\frac{dx}{dt} &=& \frac{dx}{d\tau}\frac{d\tau}{dt} = \frac{dx}{d\tau}\, f'(t) = \frac{dx}{d\tau}\,i\omega\,\tau, \\[4mm]
\frac{d^2x}{dt^2} &=& \frac{d}{dt}\left(\frac{dx}{dt}\right) =\frac{d}{dt} \left(\frac{dx}{d\tau}\,f'(t) \right) = \nonumber \\[2mm]
                              &=& f''(t)\,\frac{dx}{d\tau}+f'(t)^2 \, \frac{d^2x}{d\tau^2} =  \nonumber \\[2mm]
                              &=& -\omega^2\tau\,\frac{dx}{d\tau} -\omega^2\tau^2 \, \frac{d^2x}{d\tau^2}. \label{2odet2}
\end{eqnarray}

\noindent
Substituting this in the 2ODE (\ref{dvpeq}) we obtain:
\begin{equation}
\label{dvpeqt}
\frac{d^2x}{d\tau^2} = \frac{ \left(m\,\tau^2(x^2-1) -\omega^2\tau^2\right) \frac{dx}{d\tau} +b\,x^3\,\tau + a\,x\,\tau+f(\tau^2+1)}{\omega^2\,\tau^3}.
\end{equation}

\bigskip

\noindent
Applying the Algorithm \ref{as1} to the 2ODE (\ref{dvpeqt}) we obtain the following $S$-function
\begin{equation}
\label{sfuncdvp}
S = {\frac {-m{x}^{2}+k}{{\omega}^{2}\tau}},
\end{equation}
where the parameter $k$ was introduced (by the method) in the search for a rational $S$-function associated with the 2ODE (\ref{dvpeqt}). The present solution was found accompanied by the following relation between the 2ODE parameters:
\begin{equation}
\label{relatpar}
\left\{ a=\frac{k\,\left( k-m \right)}{\omega^2},\,b=-\frac{1}{3}\,\frac{m \left(k-m \right) }{\omega^2},\,f=f,\,m=m,\,\omega=\omega \right\}.
\end{equation}

\noindent
Applying the relation (\ref{relatpar}) to the 2ODE (\ref{dvpeqt}) we obtain
\begin{equation}
\label{dvpeqt2}
\frac{d^2x}{d\tau^2} = \frac{ \left(m \tau^2(x^2-1)\! -\!\omega^2\tau^2\right)\! \frac{dx}{d\tau} -\frac{m \left(k-m \right) }{3\omega^2}x^3\tau + \frac{k\,\left( k-m \right)}{\omega^2}x\tau+f(\tau^2+1)}{\omega^2\,\tau^3}.
\end{equation}

\noindent
Continuing the application of Algorithm \ref{as1} to the 2ODE (\ref{dvpeqt2}) we will obtain:
\begin{itemize}
\item The 1ODE associated is given by
\begin{equation}
\label{1odeass}
\frac{dv}{dx} = {\frac {-m{x}^{2}+k}{{\omega}^{2}\tau}},\,\,(v \equiv \frac{dx}{d\tau}).
\end{equation}
\item The solution of the 1ODE associated is
\begin{equation}
\label{hfunc}
H(x,v)=\frac{-m\,x^3+3\,\omega^2\,\tau\,v+3\,k\,x}{3\,\omega^2\,\tau}=C.
\end{equation}
\item The first integral is
\begin{equation}
\label{firstI}
I(\tau,x,v)=\left(v\,\tau-\frac{m\,x^3}{3\omega^2}+\frac{k\,x}{\omega^2}+\frac {f\,\tau}{-\omega^2+k-m}+\frac{f}{\tau(\omega^2+k-m)}\right)\tau^{\frac{m-k}{\omega^2}}.
\end{equation}
\end{itemize}

\bigskip

Turning back to the original variables, we finally have
 \begin{equation}
\label{firstIorig}
I(t,x,\dot{x})=\left(\frac{k\,x}{\omega^2}-i\frac{3\dot{x}+\mu\,x^3}{3\omega}+\frac {f\,{\rm e}^{i\,\omega\,t}}{k-\omega^2-i\,\omega\,\mu}+\frac{f\,{\rm e}^{-i\,\omega\,t}}{k+\omega^2-i\,\omega\,\mu}\right){\rm e}^{-\left(\mu+i\frac{k+\omega^2}{\omega}\right)t}.
\end{equation}
%

%%%%%%%%%%%%%%%%%%%%%%%%%%%%%%%%%%%%%
\section{Conclusion}
\label{conclusion}

Since the introduction of the $S$-function \cite{Nosjpa2001}, we have been finding new applications and developments for it (other people as well, the first ones were \cite{royal}). Recently, we have developed a procedure to algorithmically search for first order Liouvillian differential invariants for 2ODEs \cite{Noscpcnv2016,Nosarxiv2017}.

Here, we introduce a new method that uses the $S$-function in an algorithm to deal with (a class of) 2ODEs presenting functions. The method introduced in \cite{Noscpcnv2016,Nosarxiv2017} deal with rational 2ODEs. We illustrate the importance of this development by tackling the forced Duffing-Van der Pol oscillator.

The method allows for an integrability analysis and we have managed to find a set of values for the parameters and find a class of first order invariants for this well known problem.

We hope to have shown to the reader that our method, hereby presented, has the potential to find new solutions to very interesting problems, such as the one we have presented.

%%%%%%%%%%%%%%%%%%%%%%%%%%%%%%%%%%%%%

\section{References}
\label{refere}

\end{document}